\documentclass{llncs}

\usepackage{booktabs} 

\usepackage{graphicx}
\usepackage{wrapfig,picins}

\usepackage{amsmath}
\usepackage{float}
\graphicspath{ {images/} }
\usepackage{color}
\usepackage{url}
\usepackage{cite}
\usepackage{booktabs} 
\usepackage{subfigure}

\usepackage{amsmath}

\DeclareGraphicsExtensions{.pdf,.png,.jpg}

\usepackage[english]{babel}

\begin{document}

\frontmatter          
\pagestyle{empty}  

\title{Skin Cancer Classification using Inception Network and Transfer Learning}
 \author{Priscilla Benedetti\inst{1} $^{ORCID:0000-0002-3029-0681}$
 \newline Damiano Perri\inst{2} $^{ORCID: 0000-0001-6815-6659}$
 \newline Marco Simonetti\inst{2} $^{ORCID: 0000-0003-2923-5519}$
 \newline Osvaldo Gervasi\inst{3} $^{ORCID: 0000-0003-4327-520X}$ 
 \newline Gianluca Reali\inst{1} $^{ORCID: 0000-0002-8567-5917}$
 \newline Mauro Femminella\inst{1} $^{ORCID: 0000-0002-6695-5956}$
 }
\institute{University of Perugia, Dept. of Engineering, Perugia, Italy\and
University of Florence, Dept. of Mathematics and Computer Science, Florence, Italy\and University of Perugia, Dept. of Mathematics and Computer Science, Perugia, Italy
}

\maketitle

\begin{abstract}
Medical data classification is typically a challenging task due to imbalance between classes. In this paper, we propose an approach to classify dermatoscopic images from HAM10000 (Human Against Machine with 10000 training images) dataset, consisting of seven imbalanced types of skin lesions, with good precision and low resources requirements. Classification is done by using a pretrained convolutional neural network. We evaluate the accuracy and performance of the proposal and illustrate possible extensions.
\end{abstract}

\keywords{Machine Learning, Convolutional Neural Network, Keras, TensorFlow}

\section{Introduction}
Training of neural networks for automated diagnosis of pigmented skin lesions can be a difficult process due to the small size and lack of diversity of available datasets of dermatoscopic images. The HAM10000 (“Human Against Machine with 10000 training images”) dataset is a collection of dermatoscopic images from different populations acquired and stored by different modalities. We used the benchmark dataset, with a small number of images and a strong imbalance among the 7 different types of lesions, to prove the validity of our approach, which is characterized by good results and light usage of resources. 

Exploiting a highly engineered convolutional neural network with transfer learning, customized data augmentation and a non-adaptive optimization algorithm, we show the possibility of obtaining a final model able to precisely recognize multiple categories, although scarcely represented in the dataset. The whole training process has a limited impact on computational resources, requiring no more than 20 GB of RAM space. The rest of paper is structured as follows: Section 2 describes the related work in the field of medical images processing. Section 3 illustrates the dataset of interest. Section 4 gives an overview of the model architecture. Section 5 includes the training process and shows experimental results. Finally, some final comments and future research directions are reported in Section 6.  

\section{Related work}
Processing of biomedical images has always been a field strongly beaten by CNN pioneers. The first related papers date back to 1991 \cite{Zhang:91}, with a strong impulse in the following years in the search for methods for automating the classification of pathologies and related diagnosis \cite{doi:10.1118/1.597177, art:2}.\newline
Nowadays, almost thirty years later, reliability of networks reached a rather high level, as well as intrinsic complexity. This reliability allowed a wide diffusion of the approach of subjecting diagnostic images to automatic classification systems, from evolutionary algorithms \cite{CHEN2008337, 10.1145/1569901.1570216, art:1, art:3} to deep networks \cite{pang2017novel, zhou2017fine, art:4, art:5}, being them either convolutive or not.
Even in the medical sector of dermatology, automatic image recognition and classification was used for decades to detect tumor skin lesions \cite{lau2009automatically, dorj2018skin}.

Recent and promising research has highlighted the possibility that properly trained machines can exceed the human recognition and classification capability to recognize skin cancers. The scores obtained are very encouraging \cite{maron2019systematic} and we are confident that in the near future the recognition capacity of these forms of pathologies will become almost total.

Today CNNs are used for image feature extraction.
Features are used for image classification\cite{bocca2019, cnnDamianoDividiti2019}.
\section{The Dataset}
\label{sec:dataset}
Dermatoscopy is often used to get better diagnoses of pigmented skin lesions, either benign or malignant.
With dermatoscopic images is also possible to train artificial neural networks to recognize pigmented skin lesions automatically. Nevertheless, training requires the usage of a large number of samples, although the number of high quality images with reliable labels is either limited or restricted to only a few classes of diseases, often unbalanced.

Due to these limitations, some previous research activities focused on melanocytic lesions (in order to differentiate between a benign and malignant sample) and disregarded non-melanocytic pigmented lesions, even if very common. In order to boost research on automated diagnosis of dermatoscopic images, HAM10000 has been providing the participant of the ISIC 2018 classification challenge, hosted by the annual MICCAI conference in Granada, Spain \cite{Tschandl}, specific images. 

The set of 10015 8-bit RGB color images were collected in 20 years from  populations from two different sites, specifically the Department of Dermatology of the Medical University of Vienna, and the skin cancer practice of Cliff Rosendahl in Queensland. Relevant cases include a representative collection of all important diagnostic categories of pigmented lesions\cite{Tschandl}:
\begin{itemize}
    \item \textbf{akiec: }Actinic Keratoses and Intraepithelial Carcinoma, common noninvasive variants of squamous cell carcinoma that can be treated locally without surgery. \textbf{[327 images]}
    \item \textbf{bcc: }Basal cell carcinoma, a cancer that rarely metastasizes but grows destructively if untreated. \textbf{[514 images]}
    \item \textbf{bkl: }Generic label that includes seborrheic keratoses, solar lentigo and lichen-planus like keratoses (LPLK), which corresponds to a seborrheic keratosis or a solar lentigo with inflammation and regression, often mistaken for melanoma. \textbf{[1099 images]}
    \item \textbf{df: }Dermatofibroma, a benign skin lesion. \textbf{[115 images]}
    \item \textbf{nv: }Melanocytic nevi are benign neoplasms of melanocytes. \textbf{[6705 images]}
    \item \textbf{mel: }Melanoma, if diagnosed in an early stage, it can be cured by simple surgical excision. \textbf{[1113 images]}
    \item \textbf{vasc: }Vascular skin lesions in the dataset range from cherry angiomas to angiokeratomas and pyogenic granulomas. \textbf{[142 images]}
\end{itemize}
More than 50\% of lesions are confirmed through histopathology (histo), the ground truth for the rest of the cases is either follow-up examination (followup), expert consensus (consensus), or confirmation by in-vivo confocal microscopy (confocal). Other features in the individual dataset include age, gender  and  body-site of lesion (localization)\cite{Tschandl}.

\section{Model Architecture}
Since their first appearance in Le Cun \textit{et al.} publication \cite{leCun1998}, Convolutional Neural Networks (CNN) have been widely applied to data that have a known, grid-like structure. Possible set of interests are time-series data, which can be modeled as a 1D grid taking samples at regular time intervals, and image data, which can be thought of as a 2D grid of pixels. The foundational layer of a convolutional network consists of three stages. In the first stage, the layer performs several parallel convolutions to produce a set of linear activations. In the second stage, each convolution output is run through a nonlinear activation function, such as the rectified linear activation function (ReLU). In the third stage, a pooling function is used to modify the output of the layer further \cite{Goodfellow}. The max pooling operation, used in this work, reports the maximum output within a rectangular neighborhood.
Since the objective images of skin lesions present great variation in size, we decided to use a network made by inception modules, which make use of filters of different size operating at the same level. This usage of wide modules with multiple cheap convolutional operations entails a reduced computational complexity with respect to a deep network with large convolutional layers. 
In the specific model we used, based on Inception-ResNet-v2, another point of speed improvement is the introduction of residual connections, which replace pooling operations within the main inception modules. However, the previously mentioned max pooling operations are still present in the reduction blocks. The structure of the network used in this work is shown on Fig.1
\begin{figure}[H]
\includegraphics[width=0.3\textwidth]{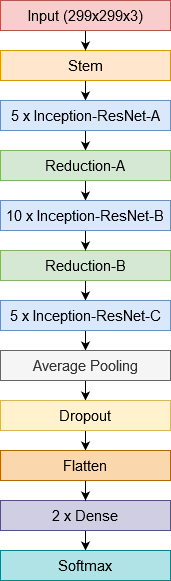}
\centering
\caption{General schema for the used network, an Inception-ResNet-v2 architecture with the addition of a flattening layer, 2 fully-connected layers and a final softmax activation.}
\end{figure}
The original Inception-ResNet-v2 architecture \cite{szegedy2016inceptionv4} has a stem block consisting of the concatenation of multiple convolutional and pooling layers, while Inception-ResNet blocks (A, B and C) contain a set of convolutional filters with an average pooling layer. As prevously mentioned, reduction blocks (A, B) replace the average pooling operation with a max pooling one. This structure has been extended with a final module consisting of a flattening step, two fully-connected layers of 64 units each, and the softmax classifier. The overall module is trainable on a single GPU with reduced memory consumption.
\section{Training process and experimental results}
This work consists of two training rounds, after a step of data processing in order to deal with the strong imbalance of the dataset:
\begin{itemize}
    \item A first classification training process using class weights.
    \item Rollback of previous obtained best model to improve classification performance with a second training phase.
\end{itemize}
\subsection{Data Processing}
In the first stage of data processing, after the creation of a new column with a more readable definition of labels, each class was translated into a numerical code using \textit{pandas.Categorical.codes}. Afterwards, missing values in "age" column was filled with column mean value. Fig.2 and Fig.3 show the HAM10000 data distribution.
\begin{figure}[H]
\includegraphics[width=0.8\textwidth]{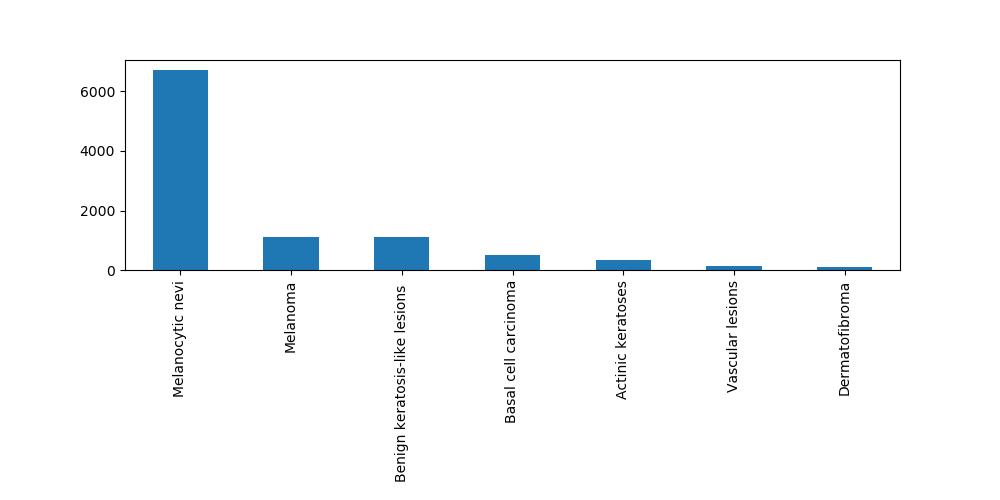}
\centering
\caption{Plotting the frequency of each class, the imbalance between Melanocytic Nevi and the rest of the possible categories is manifest.}
\end{figure}
\begin{figure}[H]
\includegraphics[width=0.8\textwidth]{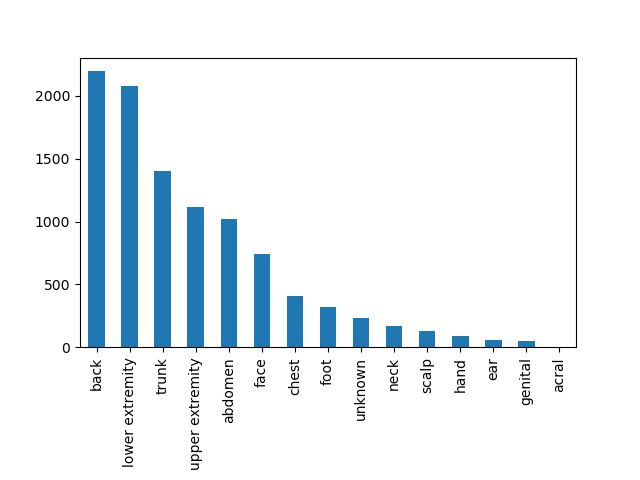}
\centering
\caption{Plot of samples' body locations.}
\end{figure}
Finally, images are loaded and resized from $450 \times $600 to $299 \times $299 in order to be correctly processed by the network. After a normalization step on RGB arrays, we split the dataset into a training and validation set with 80:20 ratio.
\begin{figure}[H]
\includegraphics[width=\textwidth]{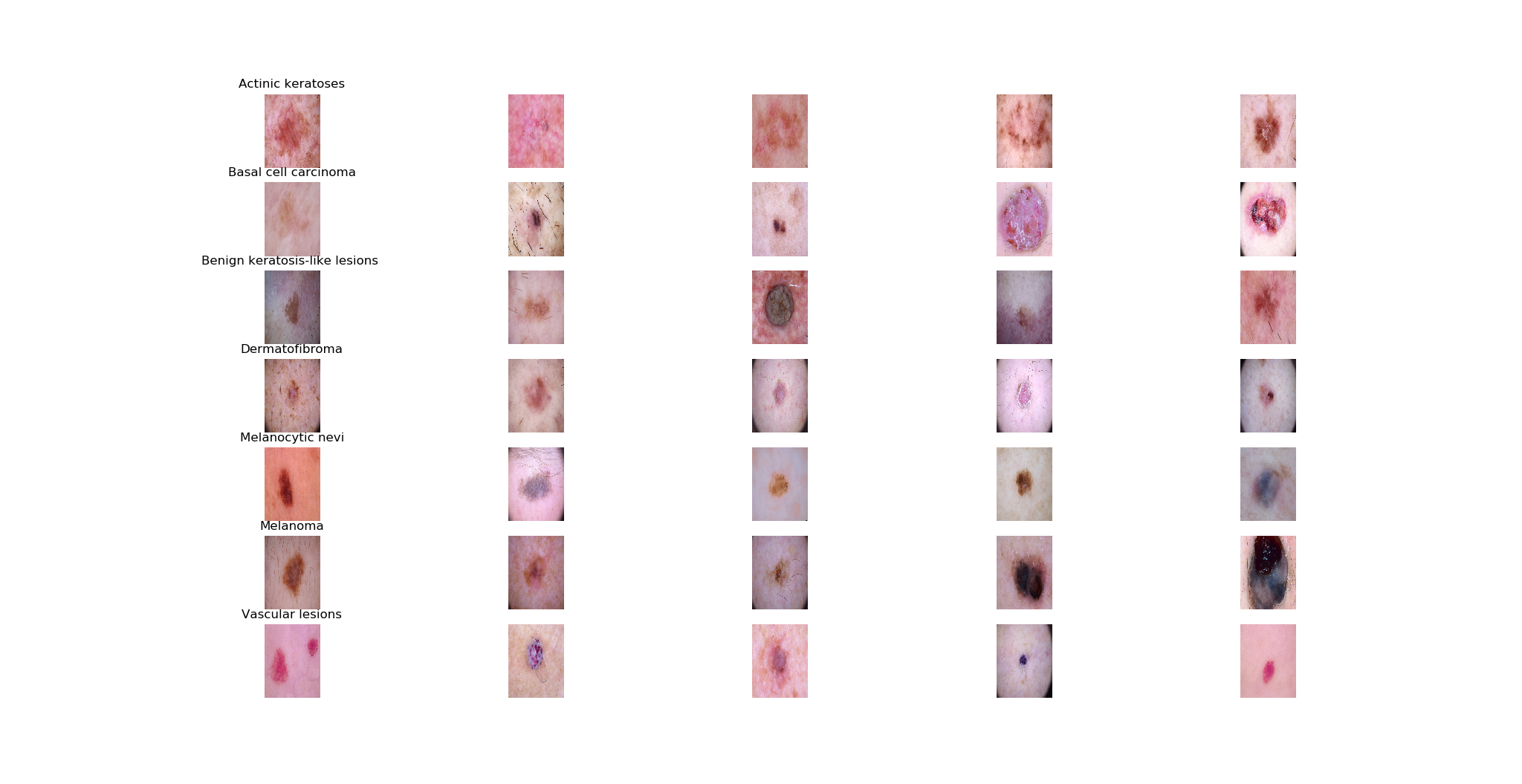}
\centering
\caption{Some HAM10000 images resized with OpenCV.}
\end{figure}
In order to re-balance the dataset, we chose to shrink the amount of images for each class to an equal maximum dimension of 450 samples. This significant decrease of available images is then mitigated by applying a step of data augmentation. Training set expansion is made by altering images with small transformations to reproduce some variations, such as horizontal flips, vertical flips, translations, rotations and shearing transformations.
\subsection{Baseline}
Due to the limited number of samples for the training process, we decided to take advantage of transfer learning, utilizing Inception-ResNet-v2 pre-trained on ImageNet\cite{Krizhevsky2012} and Tensorflow, a deep learning framework developed by Google, for fine-tuning of the last 40 layers.
Keras library offers a wide range of optimizers: Adaptive optimization methods such as AdaGrad, RMSProp, and Adam are widely used for deep neural networks training due to their fast convergence times. However, as described in \cite{wilson2017}, when the number of parameters exceeds the number of data points these optimizers often determine a worse generalization capability compared with non-adaptive methods. In this work we used a stochastic gradient descent optimizer (SGD), with learning rate set to 0.0006 and usage of momentum and Nesterov Accelerated Gradient in order to adapt updates to the slope of the loss function (categorical crossentropy) and speed up the training process.
The total number of epochs was set to 100, using a small batch size of 10. A set of class weight was introduced in the training process to get more emphasis on minority class recognition. A maximum patience of 15 epochs was set to the early stopping callback in order to mitigate the overfitting visible in Fig.5, which shows the history of training and validation process.
\begin{figure}[H]
\includegraphics[width=0.6\textwidth]{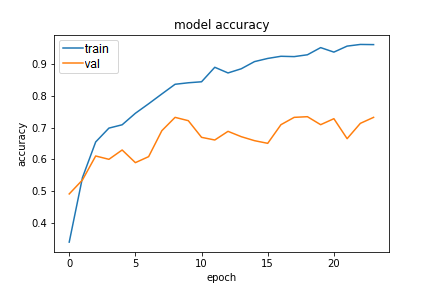}
\centering
\caption{Accuracy and Loss for each epoch.}
\end{figure}
Finally, the model achieves an accuracy of 73.4\% on the validation set, using weights from the best epoch. Fig.6 shows the confusion matrix for the model on the validation set. 
\begin{figure}[H]
\includegraphics[width=0.6\textwidth]{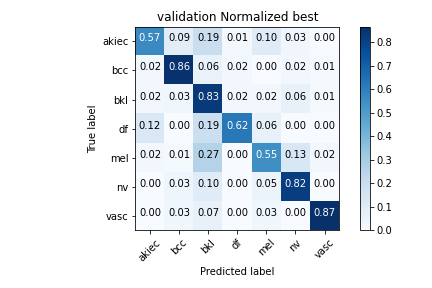}
\centering
\caption{Two of the minority classes, Actinic  Keratoses (akiec) and Dermatofibroma (df), are not properly recognized. Melanoma (mel) is often mistaken for generic keratoses (bkl), as already mentioned in \ref{sec:dataset}}
\end{figure}
\subsection{Resuming training from the best epoch: }
In order to improve classification performance, specially on minority classes, we loaded the best model obtained in the first round to extend the training phase and explore other potential local minimum points of the loss function, by using an additional amount of 20 epochs. This second step led to an enhancement in overall predictions, reaching the maximum accuracy value of 78.9\%. 
\begin{figure}[H]
\includegraphics[width=0.6\textwidth]{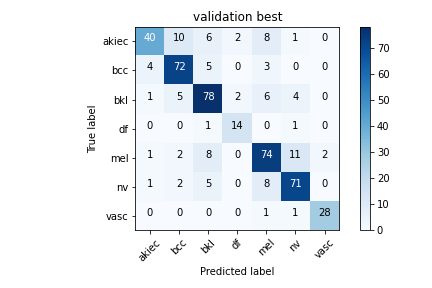}
\centering
\caption{Final confusion matrix.}
\end{figure}
\begin{figure}[H]
\includegraphics[width=0.6\textwidth]{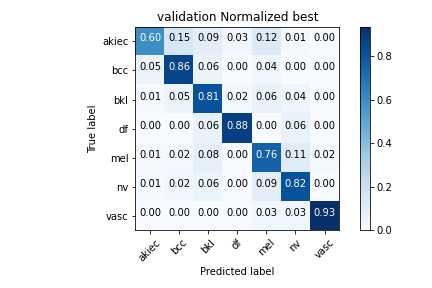}
\centering
\caption{Normalized final confusion matrix.}
\end{figure}
Fig.7 shows the normalized confusion matrix on the validation set for the final fine-tuned model. In this case, 6 out of 7 categories are classified with a total ratio of True Positives higher than 75\%, even in presence of extremely limited sample set, as  vascular lesions (vasc), 30 samples, and dermatofibroma (df), 16 samples. The whole process of training has required less than four hours on Google Colab cloud's GPU, for an overall RAM utilization below 20 GB.
\section{Conclusions and future works}
In conclusion, in this paper we investigate the possibility of obtaining improved performances in the classification of 7 significantly unbalanced different types of skin diseases, with a small amount of available images. With use of a fine-tuned deep inception network, data augmentation and class weights, the model can achieve a good final diagnostic accuracy. The described training process has a light resource usage, requiring less than 20 GB of RAM space, and it can be executed in a Google Colab notebook. For future improvements larger datasets of dermatoscopic images are needed. The model shown in this paper can be regarded as a starting point to implement a lightweight diagnostic support system for dermatologists, for example in the Web as well as through a mobile application.

%
%

\end{document}